\documentclass[twocolumn]{aastex701}

\usepackage{comment}
\usepackage{rotating}
\usepackage{threeparttablex, tablefootnote}
\usepackage[colorinlistoftodos]{todonotes}

\begin{document}

\title{A Thick Volatile Atmosphere on the Ultrahot Super-Earth TOI-561~b}

\author[0009-0008-2801-5040]{Johanna K. Teske}
\affiliation{Earth and Planets Laboratory, Carnegie Institution for Science, 5241 Broad Branch Road, NW, Washington, DC 20015, USA}
\affiliation{The Observatories of the Carnegie Institution for Science, 813 Santa Barbara St., Pasadena, CA 91101, USA}
\email{jteske@carnegiescience.edu}

\author[0000-0003-0354-0187]{Nicole L. Wallack}
\affiliation{Earth and Planets Laboratory, Carnegie Institution for Science, 5241 Broad Branch Road, NW, Washington, DC 20015, USA}
\email{nwallack@carnegiescience.edu}

\author[0000-0002-4487-5533]{Anjali A. A. Piette}
\affiliation{School of Physics \& Astronomy, The University of Birmingham; Edgbaston, Birmingham, B15 2TT, UK}
\email{a.a.a.piette@bham.ac.uk}

\author[0000-0003-4987-6591]{Lisa Dang}
\affiliation{Waterloo Centre for Astrophysics and Department of Physics and Astronomy, University of Waterloo; Waterloo, Ontario, Canada N2L 3G1}
\affiliation{Département de physique and Institut Trottier de recherche sur les exoplanètes, Université de Montréal, C.P. 6128, Succ. Centre-ville, Montréal, H3C 3J7, Québec, Canada}
\email{lisa.dang@uwaterloo.ca}

\author[0000-0002-3286-7683]{Tim Lichtenberg}
\affiliation{Kapteyn Astronomical Institute, University of Groningen; Groningen, The Netherlands}
\email{tim.lichtenberg@rug.nl}

\author[0000-0002-9479-2744]{Mykhaylo Plotnykov}
\affiliation{Department of Physics, University of Toronto, 27 King's College Cir, Toronto, ON M5S, Canada}
\email{mykhaylo.plotnykov@mail.utoronto.ca}

\author[0000-0002-5887-1197]{Raymond T. Pierrehumbert}
\affiliation{Atmospheric, Oceanic, and Planetary Physics, Department of Physics; University of Oxford, Oxford OX1 3PU, UK}
\email{raymond.pierrehumbert@physics.ox.ac.uk}

\author[0009-0009-5036-3049]{Emma Postolec}
\affiliation{Kapteyn Astronomical Institute, University of Groningen; Groningen, The Netherlands}
\email{e.n.postolec@rug.nl}

\author[0009-0003-5977-9581]{Samuel Boucher}
\affiliation{Département de physique and Institut Trottier de recherche sur les exoplanètes, Université de Montréal, C.P. 6128, Succ. Centre-ville, Montréal, H3C 3J7, Québec, Canada}
\email{samuel.boucher.3@umontreal.ca}

\author[0009-0000-8327-2631]{Alex McGinty}
\affiliation{Atmospheric, Oceanic, and Planetary Physics, Department of Physics; University of Oxford, Oxford OX1 3PU, UK}
\email{alex.mcginty@physics.ox.ac.uk}

\author[0009-0009-6098-296X]{Bo Peng}
\affiliation{Department of Earth and Planetary Sciences, Stanford University; Stanford, CA, 94305, USA}
\affiliation{Department of Astronomy \& Astrophysics, University of Toronto; Toronto, Ontario, Canada M5S 3H4}
\email{bpengeps@stanford.edu}

\author[0000-0003-3993-4030]{Diana Valencia}
\affiliation{Department of Astronomy \& Astrophysics, University of Toronto; Toronto, Ontario, Canada M5S 3H4}
\email{diana.valencia@utoronto.ca}

\author[0000-0002-6893-522X]{Mark Hammond}
\affiliation{Atmospheric, Oceanic, and Planetary Physics, Department of Physics; University of Oxford, Oxford OX1 3PU, UK}
\email{mark.hammond@physics.ox.ac.uk}

\begin{abstract}

Ultrashort-period (USP) exoplanets -- with $R_p \leq 2~$R$_{\oplus}$ and periods $\leq$1 day -- are expected to be stripped of volatile atmospheres by intense host star irradiation, which is corroborated by their nominal bulk densities and previous eclipse observations consistent with bare-rock surfaces. However, a few USP planets appear anomalously underdense relative to an Earth-like composition, suggesting an exotic interior structure (e.g., coreless) or a volatile-rich secondary atmosphere increasing their apparent radius. Here, we present the first dayside emission spectrum of the low-density (4.3$\pm$0.4 g~cm$^{-3}$) USP planet TOI-561~b, which orbits an iron-poor, alpha-rich, $\sim$10 Gyr old thick-disk star. Our 3-5~$\mu$m JWST/NIRSpec observations demonstrate the dayside of TOI-561~b is inconsistent with a bare-rock surface at high statistical significance, suggesting instead a thick volatile envelope that is cooling the dayside to well below the $\sim$3000~K expected in the bare-rock or thin-atmosphere case. These results reject the popular hypothesis of complete atmospheric desiccation for highly irradiated exoplanets and support predictions that planetary-scale magma oceans can retain substantial reservoirs of volatiles, opening the geophysical study of ultrahot super-Earths through the lenses of their atmospheres.  

\end{abstract}


\section{Introduction}

Decades of surveying nearby stars for small dips or wobbles in their light have revealed that many host short-period planets that, based on their densities, currently lack large H/He gas envelopes (see \citealt{Wordsworth_Kreidberg2022};  \citealt{Lichtenberg2025} for reviews). 
Whether these vaguely ``rocky'' planets have substantive atmospheres is a question at the forefront of exoplanet composition and habitability studies. Overall, the predicted prevalence of atmospheres on these ``super-Earth'' exoplanets is contested, as the abundance of H-C-N-S compounds is highly sensitive to delivery and loss mechanisms throughout planetary formation \citep{Krijt2023} and evolution \citep{Lichtenberg2025}. However, the most highly irradiated super-Earths are expected to be completely depleted of volatiles through atmospheric evaporation \citep{Owen2019, 2017ApJ...843..122Z}, irrespective of initial volatile endowment \citep{Burn2024, Venturini2024}, and indeed several are observationally consistent with bare rocks \citep{Kreidberg2019,Greene2023Natur.618...39G, Zieba2023,2024ApJ...961L..44Z, 2024ApJ...975L..22W, Luque2025}. 

Ultrashort-period (USP) planets -- $R_p$$\leq$2 $R_{\oplus}$, $P$$\leq$1 day -- are the most extreme cases, and most of their bulk densities match an Earth-like rock-iron mix (\citealt{Dai2021}). These ``lava worlds'' have dayside surface temperatures exceeding the melting point of most rock-forming minerals and are usually located within the disk dust sublimation radius, where in situ formation is unlikely if not impossible. While USP planets are about as common as hot Jupiters (around $\sim$0.5\% of Sun-like stars; \citealt{Sanchis-Ojeda2014}), they do not appear to be hot Jupiter remnants, based on the mismatch between host star metallicities (\citealt{winn2017}) and kinematic ages (\citealt{Hamer&Schlaufman2020}). Migration is invoked to explain USP planets, although it could be due to interaction with other planets (\citealt{Petrovich2019,Pu&Lai2019}) or a more distant companion (\citealt{Becker&Adams2017}), headwind migration combined with FU Ori outbursts (\citealt{Becker2021}), tidal interaction (\citealt{Millholland&Spalding2020}), or a combination of these mechanisms. Determining the compositions of USP planets may help shed light on their dominant formation channel. 

Intriguingly, a small subset of USP planets have densities that are significantly lower than a canonical Earth-like composition, suggesting they are enveloped by vapor. These extremely irradiated planets seem to defy expectations of atmospheric loss -- intense atmospheric escape should rapidly strip USP exoplanets of their volatiles on $<<$ Gyr timescales, based on hydrogen escape arguments \citep{Owen2019}. One proposed explanation is that volatiles may be locked up in the planet interior \citep{Dorn&Lichtenberg2021} and that a secondary atmosphere (replenished from the interior), made of volatiles heavier than helium \citep{Piette2023, Zilinskas2023, Falco2024, Peng2024}, may have experienced inhibited escape, due to processes such as radiative cooling \citep{Chatterjee2024, Ji2025}. The volatile envelope scenario would offer the exciting opportunity to, for the first time, observationally investigate the exchange of atmospheric volatiles with the interiors in rocky exoplanets. 
Emission spectroscopy of the low-density USP exoplanet 55 Cnc e hints at the possible presence of volatiles \citep{Hu2024}, but the strongly and rapidly varying occultation depths render these observations inconclusive \citep{Patel2024}, requiring characterization of other, lower-mass USP planets to establish whether they can still host volatile envelopes.  

The USP exoplanet TOI-561~b is of principal interest due to its anomalously low density ($\rho \sim$4.3$\pm$0.4~g~cm$^{-3}$; see Table \ref{tab:sys_params}) yet extremely hot temperature ($T_{\rm{eq}}\sim$2300~K). It also orbits around a thick-disk host star that is old ($\sim$10 Gyr), iron-poor, and alpha-element-rich \citep{Lacedelli2021, Weiss2021, Brinkman2023}, making its formation environment distinct from the solar system as well as from other USP planets. Here, we present the first 3-5~$\mu$m dayside emission spectrum of TOI-561~b as measured via four secondary-eclipse observations with JWST/NIRSpec \citep{Teske2023}. In \S\ref{sec:obs}, \ref{sec:data_reduction}, and \S\ref{sec:light_curve_fitting} we describe our observations, data reduction via two pipelines, and fitting the light curves to generate an emission spectrum. We go over our theoretical analysis and interpretation of the emission spectrum in \S\ref{sec:emission}, and we provide some broader context and conclusions in \S\ref{sec:discussion_conclusions}. 

\section{Observations \label{sec:obs}}

We observed TOI-561~b with JWST/NIRSpec from 2024 May 1 11:16 UT to 2024 May 3 00:37 UT, covering four consecutive secondary eclipses, whereby the planet passes behind the star. Our observation began $\sim$3.3 hr prior to the start of the first eclipse and ended $\sim$1.9 hr after the fourth eclipse to acquire an adequate out-of-transit baseline for the first and fourth eclipses. We used the highest resolution grating (G395H, resolving power$\sim$2700), which provides spectroscopy between 2.67 and 5.14 $\mu$m across the NRS1 and NRS2 detectors, excluding a detector gap between 3.72 and 3.82 $\mu$m. The observations were taken in NIRSpec Bright Object Time Series mode with the S1600A1 slit and F290LP filter, using the SUB2048 subarray and the NRSRAPID readout mode. In total, the observation consisted of 21,228 integrations with six groups per integration (6.314 s per integration), which were split into 14 segments. The JWST data analyzed in this Letter can be found in MAST: \dataset[10.17909/3g6t-he86]{http://dx.doi.org/10.17909/3g6t-he86}. 
Here, we focus on the emission spectrum; analysis of the transmission spectrum and full phase curve of TOI-561~b will be presented in future work.

\section{Data Reduction} 
\label{sec:data_reduction}

We analyzed the NIRSpec data with two different pipelines (\texttt{ExoTiC JEDI} and \texttt{Eureka!}, see below), both of which start with the raw uncal files and proceed with the standard Stage 1 and 2 steps of the STScI \texttt{jwst} pipeline \citep{Bushouse2022}, which includes corrections for linearity, dark current, and saturation. We also perform a group-level background subtraction, to account for 1/$f$ noise, and either mask or replace pixels with poor-data-quality flags. The two pipelines differ in how they trace and extract the stellar spectrum; but as shown below, they produce comparable final results, giving confidence in their robustness. None of our data are saturated and we do not divide the spectroscopic light curves by the white-light curves as was done previously for NIRCam grism data for 55 Cnc e \citep{Hu2024}.

\subsection{Eureka!}
\label{sec:eureka}
We reduced the observations using the open-source reduction pipeline \texttt{Eureka!} \citep{Bell2022}. \texttt{Eureka!} is an end-to-end pipeline for the reduction of JWST and Hubble Space Telescope data and has been used extensively in the literature for the reduction of transiting small-planet observations with JWST \citep{2024ApJ...961L..44Z,Greene2023Natur.618...39G, Zieba2023,Alderson2023,Moran2023,Wallack2024}. We used \texttt{Eureka!} in a similar manner as described in \cite{Wallack2024}, and we describe the process briefly herein. 

\texttt{Eureka!} acts as a wrapper around the \texttt{jwst} pipeline for Stage 1 and Stage 2. We used \texttt{Eureka!} version 0.10 and \texttt{jwst} version 1.11.4 \citep{Bushouse2023}, utilizing the default \texttt{jwst} stages, with the exception of the inclusion of the custom \texttt{Eureka!} group-level background subtraction that is utilized in Stage 1. Stage 3 and Stage 4 provide additional calibrations, such as outlier rejection and trace rectification, and extract the time series of stellar spectra, as well as the white-light and spectroscopic light curves, and allow for a number of different reduction optimizations to be done. 

We tested different combinations of these parameters (specifically, the aperture width for extraction, the background aperture width, additional background subtractions, and the sigma threshold for the outlier rejection during optimal extraction) in order to determine the most optimal parameters for the reduction. To only analyze the secondary eclipses and transits without the overprinted effects of any phase modulation, we separated each observational segment into transits and eclipses, including adequate pre- and post-transit baselines (rounded to the nearest observation segment), but not the entire phase curve. This resulted in four separate eclipse observations (of varying numbers of segments) and three different transit observations. We optimized the reduction parameters of each eclipse separately, to try to limit the effects of any long-term time variability. This means that we treat each individual eclipse separately and do not fit for an overall systematic noise model or phase curve. We optimize each parameter by generating a white-light curve with each set of reduction parameters and taking the combination that minimizes the median absolute deviation. We consider extraction apertures from 4 to 8 pixel half-widths, and background apertures of 8-11 pixels. We also optimize whether we do a full-frame or column-by-column additional background subtraction and the sigma threshold for the outlier rejection during the optimal extraction. We pick the combination of parameters that produces the minimum of the median absolute deviation as a proxy for minimizing the scatter in the white-light curves. We find that all four eclipses in both detectors favor 4 pixel half-widths for the extraction aperture and 8 pixel background apertures. All NRS1 and NRS2 eclipses favor an additional column-by-column background subtraction over a full-frame background subtraction. All NRS1 eclipses favor a sigma threshold of 60, as does Eclipse 1 for NRS2. The three other NRS2 eclipses favor a sigma threshold of 10.

\subsection{ExoTiC JEDI}
\label{sec:jedi}

We also reduced the data using the Exoplanet Timeseries Characterisation-JWST Extraction and Diagnostic Investigator (\texttt{ExoTiC JEDI}) pipeline \citep{Alderson2022JEDI}, which has been used and explained in numerous other  publications of JWST NIRSpec/G395H data \citep{Grant2023, Alderson2024, Alderson2025, Wallack2024, Kirk2025}. As in those cases, we treated the NRS1 and NRS2 detectors independently. 

Starting from the raw \texttt{uncal} files, \texttt{ExoTiC JEDI} begins Stages 1 and 2 operating with a modified version of the STScI \texttt{jwst} pipeline (v1.13.4, context map 1100; \citealt{Bushouse2022}). In Stage 1, the pipeline corrects for linearity, dark current, and saturation, and uses a jump detection threshold of 15$\sigma$. We perform a custom bias subtraction by creating a median of all of the first groups across integrations in the time series and subtracting this median image from all groups. We also perform a custom ``destriping'' step, to remove the background 1/$f$ noise, in which we mask out the spectral trace within 15 times the standard deviation of the point-spread function along the dispersion direction for each integration, then subtract the median pixel value from each column at the group level. \texttt{ExoTiC JEDI} uses the standard ramp-fitting step from Stage 1 of the \texttt{jwst} pipeline and the standard Stage 2 steps to generate a 2D wavelength map. In Stage 3, \texttt{ExoTiC JEDI} identifies any pixels with data-quality flags ``do not use'', ``saturated'', ``non science'', ``dead'', ``hot'', ``low quantum efficiency'', and ``no gain value'' and replaces them with the median of the nearest four pixels. Additionally, any pixels that are outliers at the 20$\sigma$ level or more temporally are replaced by the median of the nearest 10 integrations, and outliers at the 6$\sigma$ level or more spatially are replaced with the median of the nearest four pixels. A second round of removing 1/$f$ and background noise is performed via subtracting the median unilluminated (outside the spectral trace) pixel value from each column in each integration. Finally, to extract the 1D stellar spectra, \texttt{ExoTiC JEDI} generates a spectral trace by fitting a Gaussian to each column, fitting fourth-order polynomials to both the trace center and width, and then smoothing with a median filter 5 pixels in width. The aperture region is then taken to be 5 times the FWHM of the trace, or approximately 8 pixels across, and intrapixel extraction is applied (accounting for partial pixels). In addition to the 1D stellar spectra, \texttt{ExoTiC JEDI} produces $x$- and $y$-position shifts as a function of time, calculated by cross-correlating the spectra; these can be used for detrending systematic noise in the light curves. The wavelength solution for the 1D spectrum is extracted from the wavelength map using the same trace positions as the spectral extraction.

\section{Light-curve Fitting} 
\label{sec:light_curve_fitting}

To generate the emission spectrum of TOI-561~b, we applied the same fitting procedure to each of the reductions. In the case of \texttt{ExoTiC JEDI}, we first separated out in time the four secondary eclipses using the different observing segments, after excluding segments close to or containing transits; this was already done in our nominal \texttt{Eureka!} reduction, as described above. We fit the extracted white-light curves (one each for NRS1 and NRS2) with a combined astrophysical model (using BATMAN; \citealt{Kreidberg2015}), an error inflation term that we add in quadrature with the measured errors, and instrumental noise model, to constrain the best-fit time of eclipse and eclipse depth. This error inflation term is meant to account for any underestimates of the pipeline determined per point measurement uncertainties of the time series, but it does not necessarily capture time-correlated noise. The instrumental noise model is of the form 
\begin{equation} \label{eq:systematics}
      S=p_1 + p_2 \times T + p_3 \times X + p_4 \times Y
\end{equation}

where $T$ is the array of times, $X$ is the array of $x$ positions of the trace on the detector, $Y$ is the array of $y$ positions of the trace on the detector, and $p_N$ is a free parameter. This systematic noise model is fit individually for each eclipse. For the astrophysical model, we fix all of the astrophysical parameters, except for the time of eclipse and $F_p$/$F_*$ to the values given in Table \ref{tab:sys_params}. We first fit the four NRS1 white-light eclipses together and separately fit the four NRS2 white-light eclipses together, assuming a common eclipse time and eclipse depth for each of the four visits within a detector. We then fixed the time of eclipse for all spectroscopic bins within a detector to the median of the Markov Chain Monte Carlo (MCMC) chain from the fit to the white-light curve for that detector, allowing the $F_p$/$F_*$ to vary in each bin and fitting the systematic model and error inflation term.

\begin{figure*}
    \centering
        \includegraphics[width=1\textwidth]{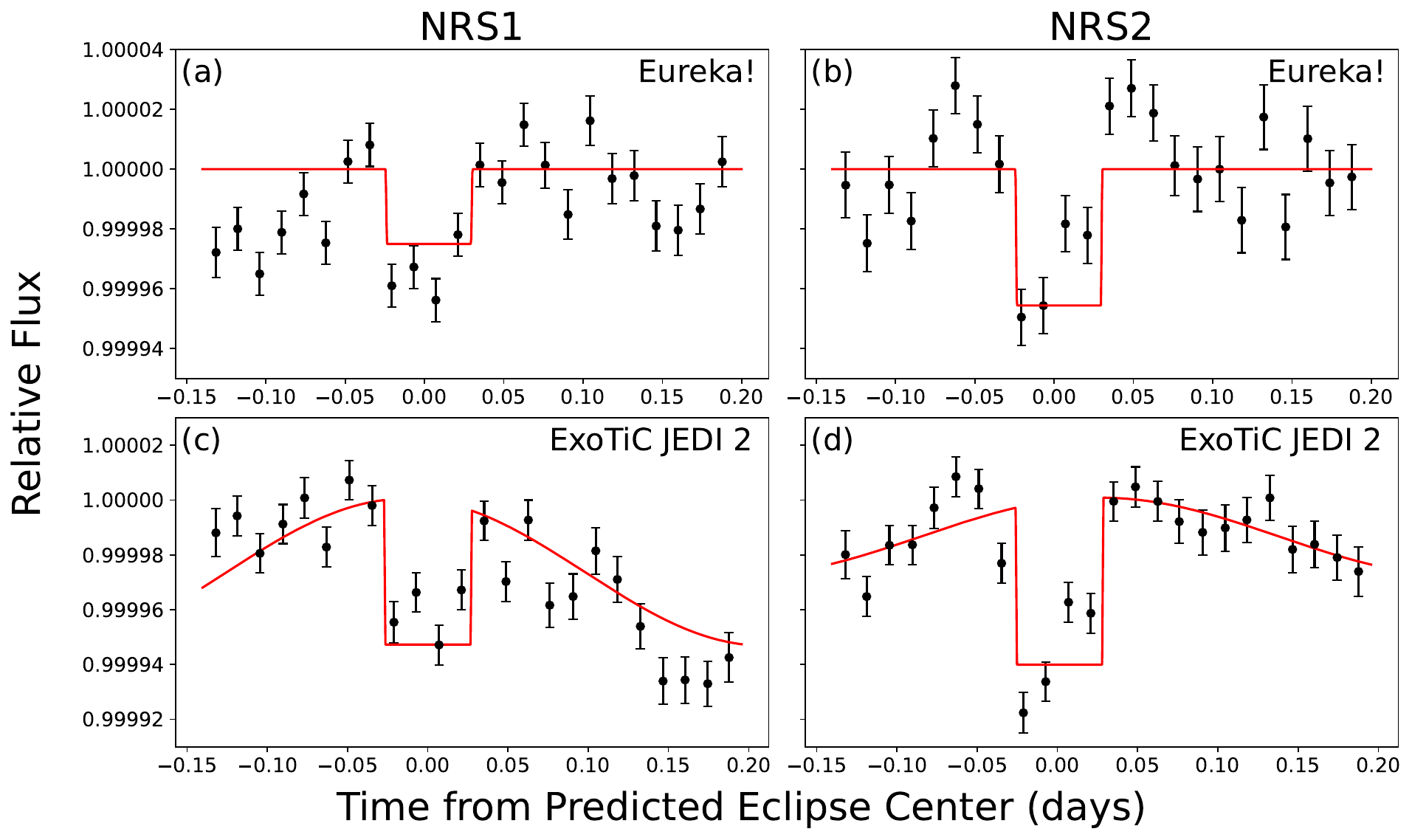}
    \caption{Phase-folded \texttt{Eureka!} (top) and \texttt{ExoTiC JEDI 2} (bottom) independently normalized white-light curves of secondary eclipses of TOI-561~b. We show the phase-folded eclipses with instrumental noise removed for (a,c) NRS1 (2.8627040 to 3.7143560 microns) and (b,d) NRS2 (3.8199180 to 5.0627574 microns) (see \S\ref{sec:light_curve_fitting} for details). The black points with associated errors show the 20 minute binned \texttt{Eureka!} or \texttt{ExoTiC JEDI 2} data and the red lines show the best-fit astrophysical model from the joint fit to the four eclipses. From the NRS1 and NRS2 \texttt{Eureka!} white-light curves, we measure eclipse depths of 25$\pm$4 ppm and 46$\pm$5 ppm for \texttt{Eureka!}, and 53$^{+6}_{-4}$ ppm and 60$^{+7}_{-8}$ ppm for \texttt{ExoTiC JEDI 2}.}
    \label{fig:WLC}
\end{figure*}

The best-fit parameters and associated uncertainties were determined using the affine-invariant MCMC ensemble sampler \texttt{emcee} \citep{Foreman-Mackey2013}, where walkers were initialized at the best fit from a Levenberg–Marquardt least-squares minimization. In Figure~\ref{fig:WLC}, we show the phase-folded NRS1 and NRS2 white-light curves; each point represents the error-weighted average and associated error across each time bin for all four eclipses after removing the different noise models from each eclipse. Overplotted in this figure is the model derived from jointly fitting all four eclipses for each detector. 
In Appendix \ref{sec:Appendix_additional_figures} we show the resulting joint fits to the \texttt{Eureka!} reduction for each of the four secondary eclipses (Figure \ref{fig:individual_WLC}) and the root mean square (RMS) versus bin size curves for both the white-light curves and the spectroscopic bins (Figure \ref{fig:RMS_binsize}). In Appendix \ref{sec:Appendix_correlated_noise}, we describe an estimate of the error inflation due to correlated noise in the \texttt{Eureka!} reduction.

Given the interest in variability in lava world planets \citep{2016Demory, 2023MeierValdes, Patel2024,Zilinskas2025}, we also tried fitting the eclipse white-light curves independently, versus jointly, using a similar procedure (isolating each eclipse for fitting). In these fits, we fixed the mideclipse time to the time at phase=0.5. For both NRS1 and NRS2, the error-weighted average values and the jointly fit values are consistent -- 23$\pm$4 ppm versus 25$\pm$4 ppm for NRS1 and 47$\pm$5 ppm versus 46$\pm$5 ppm for NRS2. While there is some scatter in the NRS1 independently-fit eclipse depths, we suspect this may be due to detector systematics and/or stellar variability, so we do not suggest planetary dayside emission variability at this point. Further investigation of the full time series is described below, as well as in an upcoming paper, supporting the overall conclusions presented here. In addition, these independent fits also produce relatively little residual correlated noise, further reinforcing the robustness of our derived joint-fit spectrum, despite the presence of residual correlated noise (see also Appendix \S\ref{sec:Appendix_correlated_noise}).

For the spectroscopic light curves, we divide the data into seven ($\sim$400 pixel, $\sim$0.3 micron) wavelength bins across the two detectors and again perform a joint fit across all four eclipses for each bin; for these fits, only the $F_p$/$F_s$ was left as a free parameter. The resulting measured eclipse depths are listed in Table \ref{tab:eclipse_depths}, and the depths are referred to as \texttt{Eureka!} and \texttt{ExoTiC JEDI} 1 in Figure \ref{fig:spectrum}.

To add further robustness to our result, we also tested an alternative fitting procedure on the \texttt{ExoTiC JEDI} broadband and spectroscopic light curves that allows for baseline curvature outside of eclipses and transits due to the phase variation expected from the short orbit of the planet, and found consistent results. In this analysis, we used the same detector model described by equation \ref{eq:systematics} for consistency,  and instead of solely using BATMAN \citep{Kreidberg2015} as the astrophysical signal to fit the eclipse, we combined it with a first-order sinusoidal phase-curve model to account for the curvature of the phase curve described by equation (6) in \cite{2018NatAs...2..220D}. Similarly, we fixed all the astrophysical parameters the values given in Table \ref{tab:sys_params} and only allowed the the eclipse depth, $F_p$/$F_*$, and the phase-curve parameters to vary. We note that here we also fit for the photometric uncertainty, $\sigma _F$ which is assumed to be the same for all data points. To account for the effect of time-correlated noise, we calculated the RMS of the fit residual as a function of temporal bin size and compared them to the expected trend if the residuals were Gaussian white noise, similar to Figure \ref{fig:RMS_binsize}. We then took the ratio between our residuals' RMS and the expected RMS if the residual was uncorrelated at the temporal bin size corresponding to the duration of the eclipse and inflated our eclipse depth uncertainties by this ratio. The resulting measured white-light curves are shown in Figure \ref{fig:WLC}, and the resulting emission spectrum is shown in Figure \ref{fig:spectrum} (referred to as \texttt{ExoTiC JEDI 2}), with the measured eclipse depths listed in Table \ref{tab:eclipse_depths}. 

\begin{deluxetable*}{cccc}
\tablecaption{Measured eclipse depths \label{tab:eclipse_depths} }
\tablewidth{0pt}
\tablehead{\colhead{Wavelength}&\colhead{\texttt{Eureka!} eclipse}&\colhead{\texttt{ExoTiC JEDI 1} Eclipse} &\colhead{\texttt{ExoTiC JEDI 2} Eclipse}\\
\colhead{Bin ($\mu$m)}&\colhead{Depth $F_p$/$F_s$ (ppm)}&\colhead{Depth $F_p$/$F_s$ (ppm)} &\colhead{Depth $F_p$/$F_s$ (ppm)}} 
\startdata
2.863-3.147 & 13.80$\pm$6.58& 18.13$\pm$6.54 & 41.89$^{+8.27}_{-7.06}$ \\
3.147-3.430 &32.22$\pm$6.52 &27.55$\pm$6.47 & 42.94$^{+8.35}_{-9.81}$\\
3.430-3.714 &27.61$\pm$7.00 &29.91$\pm$6.92 & 52.92$^{+7.65}_{-8.39}$\\
3.820-4.131 &26.03$\pm$7.92 &32.76$\pm$7.92 & 34.95$^{+9.35}_{-10.18}$\\
4.131-4.441 &57.02$\pm$9.80 &66.37$\pm$9.95 &75.57$^{+12.60}_{-11.08}$\\
4.441-4.752 &45.01$\pm$12.16 &61.71$\pm$12.50 & 71.79$^{16.69}_{-14.21}$\\
4.752-5.063 &51.34$\pm$15.39 &62.24$\pm$15.99 &72.54$^{+20.72}_{-23.75}$\\
\enddata 
\end{deluxetable*}


\section{Theoretical Modeling of Planet's Emission Spectrum} 
\label{sec:emission}

We model the dayside emission spectrum of TOI-561~b using two approaches: (i) Bayesian atmospheric retrievals, to place statistical constraints on the presence of spectral features and the brightness temperature of the spectrum; and (ii) self-consistent 1D atmospheric models, to determine atmospheric compositions that are (un)able to explain the data. In both approaches, we consider the \texttt{Eureka!} and \texttt{ExoTiC JEDI} reductions/fits and use the planetary and stellar parameters listed in Table \ref{tab:sys_params}. 

We first perform retrievals using a single-parameter blackbody-spectrum model to determine the best-fit brightness temperature of the data. The atmospheric retrievals are performed with the HyDRo retrieval code \citep{Piette2022, Gandhi2018, Piette2020b}, which combines a parametric model of the atmosphere with a nested sampling Bayesian parameter estimation algorithm, PYMULTINEST \citep{Skilling2006,Feroz2009,Buchner2014}. We find brightness temperatures of 1800$^{+130}_{-60}$~K for the \texttt{Eureka!} reduction, 1870$^{+80}_{-90}$~K for the \texttt{ExoTiC JEDI} reduction with the first fitting procedure, and 2150$^{+80}_{-80}$~K for the \texttt{ExoTiC JEDI} reduction with the second fitting procedure (see Figure \ref{fig:bb_temp} in Appendix \ref{sec:Appendix_additional_figures}). These are all significantly cooler than the brightness temperature expected for a bare rock (2950~K), and also cooler than the zero-albedo equilibrium temperature assuming full day-night heat redistribution (2310~K). Since the 3-5~$\mu$m range observed does not probe the peak of the planetary emission spectrum, this could be the result of: (i) a gray atmosphere with a high Bond albedo; (ii) a nongray atmosphere that is able to cool through opacity windows at shorter wavelengths, where the planetary emission peaks; or (iii) a combination of a nongray atmosphere and elevated Bond albedo, e.g., due to clouds. 

We next determine whether molecular features are detectable in the spectrum by comparing the Bayesian evidences of: (i) the single-parameter blackbody-spectrum model, and (ii) a retrieval model including the parametric temperature profile of \cite{Madhusudhan&Seager2009}, plus the abundances of H$_2$O, CO, CO$_2$ and SiO, and a parameter for the mean molecular weight of the atmosphere. The molecular opacities are calculated using line lists from the ExoMol and HITEMP line lists, as described below \citep{Rothman2010,Yurchenko2022}. We find no statistical preference for the model with molecular features, with a Bayes factor $<$1.5 between a model with molecular opacity vs the blackbody model. The more complex retrieval model prefers a solution with an almost isothermal temperature profile, illustrating the closeness of the data to a blackbody curve. We therefore determine that variations in the spectra are consistent with noise rather than chemical features. This may be influenced by the wide binning in wavelength that is required to detect the shallow eclipses.

\begin{figure*}
    \centering
        \includegraphics[width=1\textwidth]{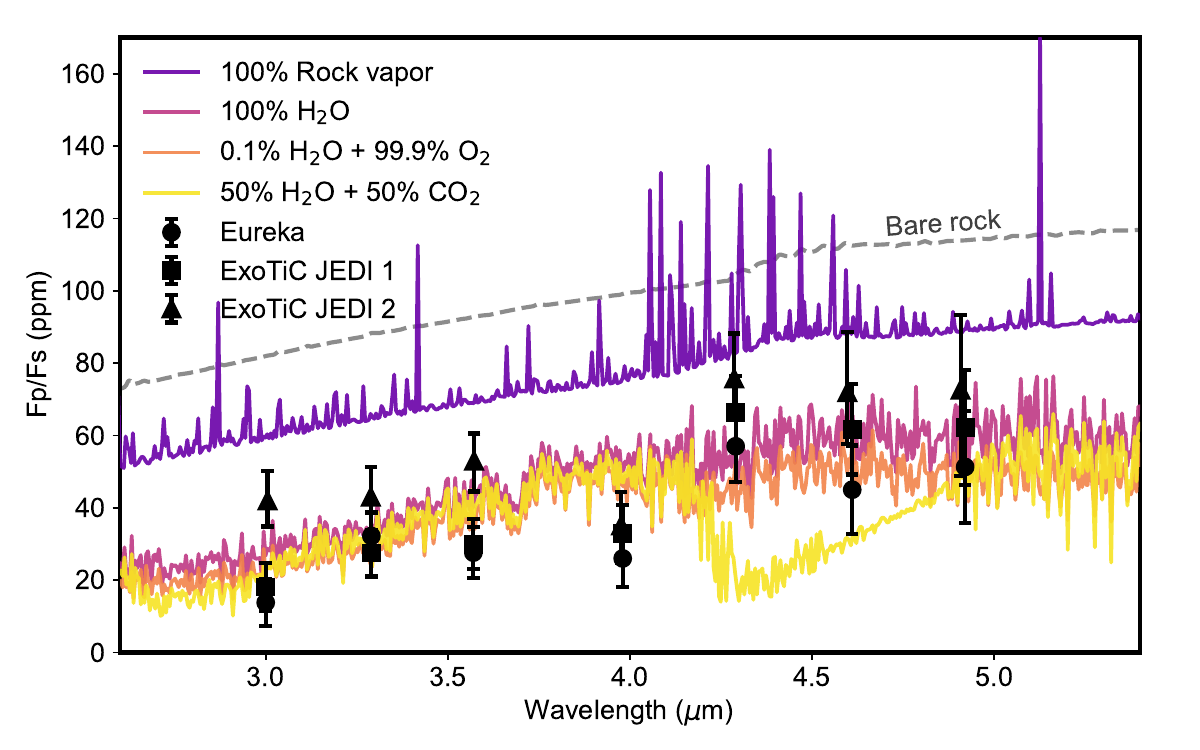}
    \caption{The JWST/NIRSpec emission spectrum of TOI-561~b is inconsistent with a zero-albedo bare-rock surface. The black symbols and error bars show the \texttt{Eureka!} (circles), \texttt{ExoTiC JEDI 1} (squares), and \texttt{ExoTiC JEDI 2} (triangles) reductions/fits. There is no offset applied between the NRS1 and NRS2 detectors (gap around 3.75~$\mu$m). The gray dashed line shows the expected emission spectrum for a bare rocky surface assuming zero Bond albedo. The colored lines show model spectra simulated using a 1D self-consistent atmospheric model and different chemical compositions (see the legend). We assume no day-night heat redistribution in the rock vapor case and efficient day-night heat redistribution in the cases without rock vapor. Volatile-rich atmospheric compositions are able to reproduce the observed brightness temperature. The points plotted here are in Table \ref{tab:eclipse_depths}.}
    \label{fig:spectrum}
\end{figure*}

Given the slight bump in the spectrum between 4 and 4.3~$\mu$m (see Fig \ref{fig:spectrum}, which we clarify lies on the same detector and is not a result of a detector offset), we also test whether a simpler model with either CO$_2$ or SiO opacity alone is statistically preferred, owing to their opacity peaks in this spectral region. We find that a retrieval model with the nonisothermal pressure-temperature profile and the abundance of CO$_2$ only or SiO only is not statistically preferred over the single-parameter blackbody-spectrum model, resulting in Bayes factors $\leq$1. 

Following these retrieval analyses, we investigate scenarios that could explain the observations using a self-consistent 1D radiative-convective equilibrium model, GENESIS \citep{Gandhi2017,Piette2020a,Piette2023}. GENESIS solves for radiative-convective, hydrostatic, and local thermodynamic equilibrium, as well as thermochemical equilibrium via coupling to FASTCHEM COND \citep{Kitzmann2024}. The atmospheric model is coupled to the VapoRock magma ocean outgassing code, which calculates the abundances of the chemical species expected to evaporate from a magma ocean surface. We consider molecular opacities due to H$_2$O \citep{Rothman2010}, CO$_2$ \citep{Rothman2010}, CO \citep{Rothman2010}, SiO \citep{Yurchenko2022}, SiO$_2$ \citep{Owens2020}, AlO \citep{Patrascu2015}, MgO \citep{Li2019}, NaO \citep{Mitev2022}, TiO \citep{McKemmish2019}, O$_2$ \citep{Gordon2017}, OH \citep{Rothman2010}, FeH \citep{Dulick2003,Bernath2020}, NaH \citep{Rivlin2015}, NaOH \citep{Owens2021}, and KOH \citep{Owens2021}. We also consider atomic and ionic opacities from Al, Ca, Fe, H, K, Mg, Na, O, Si, Ti, Ca+, and Na+ obtained from the DACE database \citep{Grimm2021}, which are calculated using the helios-k code \citep{Grimm2015, Grimm2021}, and data from the Kurucz database \citep{Kurucz1992,Kurucz2017,Kurucz2018}.

In the absence of an atmosphere, TOI-561~b would be hot enough to sustain a dayside magma ocean \citep{Lichtenberg2025} and a thin, $\sim$0.01, rock vapor atmosphere \citep{Miguel2011}. This atmosphere would nevertheless be largely optically thick, as a result of strong rock vapor opacities, notably SiO and MgO in the 3-5~$\mu$m range \citep{Piette2023,Zilinskas2022}. The brightness temperature observed would therefore be set by the atmosphere rather than the reflectivity of the magma ocean below (which is expected to be low;  \citealt{2020ApJ...898..160E} -- but see also \citealt{Kite2016}). We investigate this scenario, using the same setup and surface composition as in \cite{Piette2023} to calculate the expected thermal emission spectrum of TOI-561~b in the case of a pure-rock-vapor atmosphere (i.e., no additional volatiles present). The zero-albedo irradiation temperature of TOI-561~b, assuming no heat redistribution to the nightside, is 2750~K, at which the outgassed surface pressure is expected to be ~0.01 bar. We therefore assume a 0.01 bar surface pressure and that the atmosphere is not able to transport heat to the nightside, as expected for a thin-rock-vapor atmosphere \citep{Koll2019}. As shown in Figure \ref{fig:spectrum}, this 100\% rock vapor spectrum is clearly inconsistent with the data, with reduced chi-square values of 27.3/22.9/5.6 for the \texttt{Eureka!}/\texttt{ExoTiC JEDI 1}/\texttt{ExoTiC JEDI 2} reductions. For all three reductions, the data lie systematically below the pure-rock-vapor model, meaning that even the \texttt{ExoTiC JEDI 2} data are inconsistent with it, despite a lower reduced chi-square compared to the other reductions.

The measured dayside temperature of TOI-561~b is $\gtrsim 1000$~K lower than the maximum possible ``bare-rock'' temperature of 2950~K, which, combined with the rejection of a thin-rock-vapor atmosphere, is indicative of cooling effects from a thick atmosphere. This scenario can be explained by a combination of heat transport to the nightside, a nonzero Bond albedo, and/or nongray opacity in the atmosphere. Since the observed 3-5~$\mu$m range does not probe the peak blackbody emission of the planet ($\sim$1-2~$\mu$m), opacity windows in the optical to near-infrared allow the atmosphere to cool significantly. That is, in the case of a noninverted temperature profile (increasing temperature with depth), optical/near-infrared opacity windows would probe the deeper, hotter regions of the atmosphere, while the (more opaque) 3-5~$\mu$m range would probe the cooler, shallower regions.

We therefore investigate whether volatile-rich atmospheric compositions are consistent with the data. We consider three volatile-rich atmospheric compositions, consisting of 100\% H$_2$O, 0.1\% H$_2$O + 99.9\% O$_2$, and 50\% H$_2$O + 50\% CO$_2$. The 100\% H$_2$O and 0.1\% H$_2$O + 99.9\% O$_2$ models match the data reasonably well, with reduced chi-square values of 3.2/2.0/1.7 and 1.8/1.5/3.1, respectively, for the \texttt{Eureka!}/\texttt{ExoTiC JEDI 1}/\texttt{ExoTiC JEDI 2} reductions. The model containing CO$_2$ provides a less good fit (reduced chi-square of 4.0 or 5.3 for each reduction), owing to the CO$_2$ absorption feature at 4.5~$\mu$m, which deviates from the observed spectrum. The models plotted in Figure \ref{fig:spectrum} assume efficient day-night heat redistribution, though at these high temperatures, a lower redistribution efficiency is also possible. In the lower-efficiency case, a nonzero Bond albedo would be required to cool the dayside to the observed temperatures, e.g., as a result of silicate clouds \citep[e.g.,][]{Lee2025_silicate_clouds}.

In summary, our analysis shows that volatile-rich atmospheric compositions are able to explain the brightness temperature observed for TOI-561~b in the 3-5~$\mu$m range thanks to their opacity windows at shorter wavelengths, more efficient day-night heat redistribution, and potential to host reflective clouds. A thinner pure-rock-vapor atmosphere could also host reflective clouds, which may result in a cooler brightness temperature. However, \citet{Nguyen2024_lava_clouds} find that heat advection can compensate for the cooling effects of clouds in such an atmosphere, meaning that this scenario may still be too hot to reproduce the eclipse spectrum we observe. Furthermore, a thin-rock-vapor atmosphere would be at odds with the low observed bulk density of the planet.

\section{Discussion and Conclusions} \label{sec:discussion_conclusions}

\begin{figure*}
    \centering
        \includegraphics[width=1\textwidth]{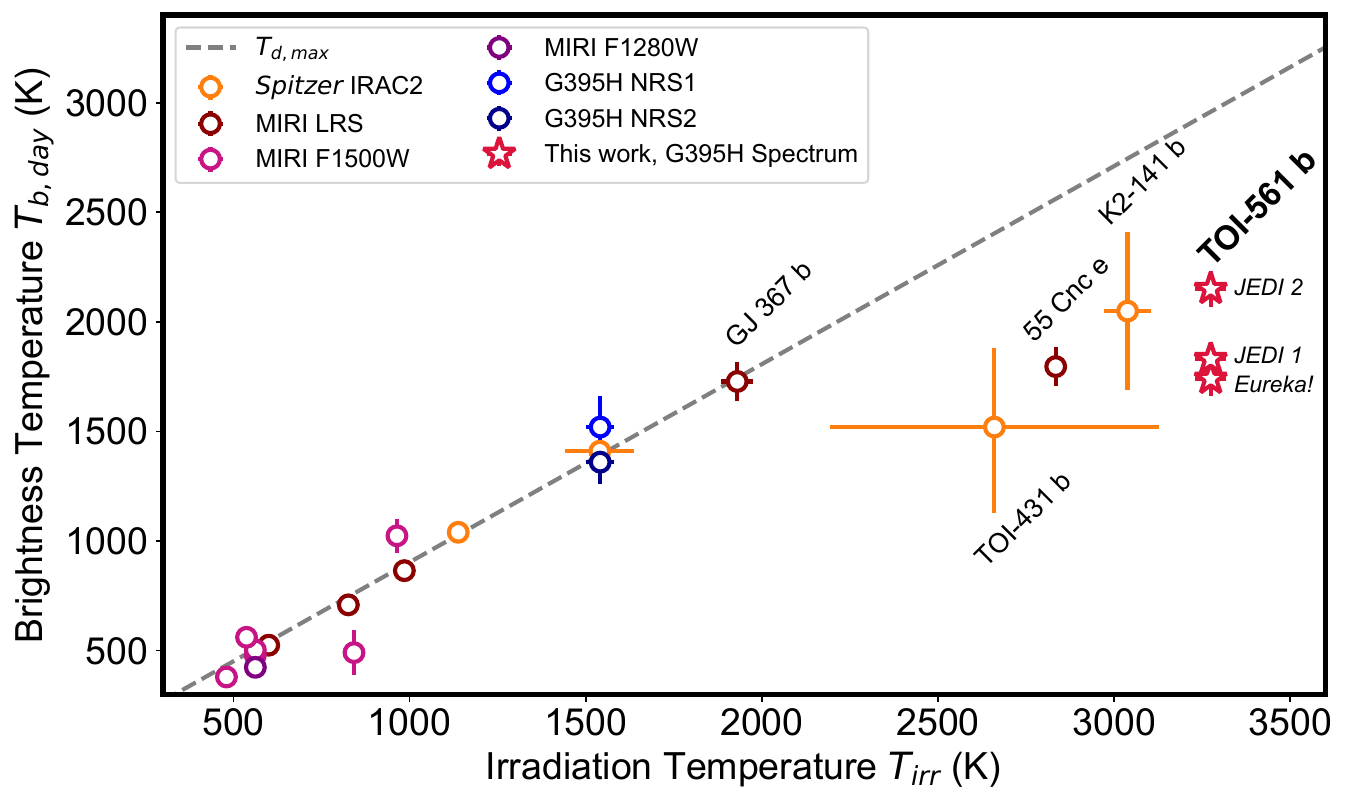}
    \caption{TOI-561~b's dayside brightness temperature presents the strongest evidence out of all rocky planets for the presence of an atmosphere. As measured from the $\sim$3-5~$\mu$m G395H emission spectrum, the brightness temperature of TOI-561~b (red star symbols, one for each data reduction/fit) is well below the irradiation temperature expected assuming a zero-albedo, zero-heat-redistribution planet (dashed gray line; see the text). The observed trend in the dayside brightness temperature of the current sample of observed rocky planets suggests that USPs with highly irradiated daysides ($T_{\rm{irr}}>$2500~K) have atmospheres (that may be rock vapor) evaporated from their molten daysides, although note that wavelength ranges are different depending on the observing mode (e.g., MIRI/LRS covers $\sim$6.5-11.5~$\mu$m). The data for this plot are provided in Table \ref{tab:brightness_temps_lit} in Appendix \ref{sec:Appendix_additional_figures}.}
    \label{fig:brightness_temp}
\end{figure*}

Our dayside brightness temperature constraint for TOI-561~b is the strongest evidence yet among USP planets for the presence of a thick atmosphere (Figure \ref{fig:brightness_temp}). To make this comparison, we computed the expected maximum dayside temperature, defined as $T_{d,\rm{max}}=T_{\rm{irr}} (1-A_{b})f^{1/4}$, as a function of irradiation temperature, $T_{\rm{irr}} = T_{\rm{*,eff}} \sqrt{R_*/a}$,  where $T_{\rm{*,eff}}$ is the effective stellar temperature of the host star and $a/R_*$ is the semi-major axis, in units of stellar radius. To obtain the maximum dayside temperature, we assume a zero bond Albedo $A_b=0$ and a heat redistribution factor $f=2/3$ corresponding to zero redistribution to the nightside \citep{Hansen2008,Cowan2011}. We then compare all dayside brightness temperature constraints of terrestrial planets published thus far (see Table \ref{tab:brightness_temps_lit} in Appendix \ref{sec:Appendix_additional_figures}) against their expected maximum dayside temperature,  as shown in Figure \ref{fig:brightness_temp} (see also \citealt{Coy2025} for a similar analysis of rocky planets around M stars).  

\begin{figure*}
    \centering
        \includegraphics[width=1\textwidth]{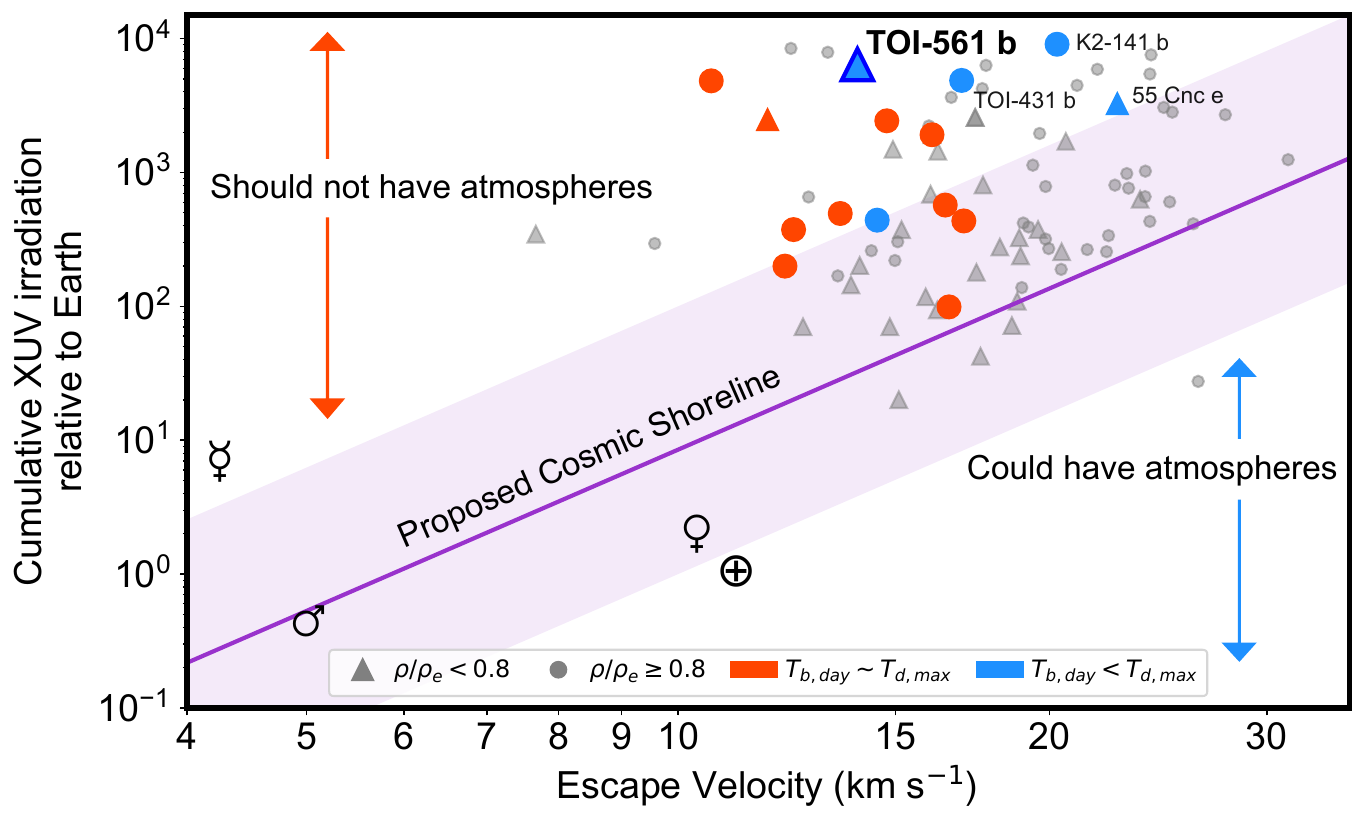}
    \caption{The ``cosmic shoreline'', which divides solar system planets into those with and without atmospheres, is not universal across exoplanetary systems. The empirical relation between the estimated cumulative XUV irradiation and escape velocity, $I_{\rm{XUV}} \propto$ \textit{v}$_{\rm{esc}}^{4}$ (solid purple line), is in line with predictions of XUV stellar radiation driving atmospheric escape in planets \citep{2017ApJ...843..122Z}; see also \cite{Chatterjee2024} and \cite{Ji2025} for more analytical predictions. The arbitrary envelope around the cosmic shoreline represents uncertainty based on factors like host star type, atmospheric composition, and initial volatile content. The cumulative XUV irradiation for each planet is estimated based on the simple scaling relation (Eq. 27) in \cite{2017ApJ...843..122Z}. Planets marked in colors have measured dayside brightness temperatures from Spitzer or JWST that are consistent with (red) or less than (blue) the maximum dayside brightness temperature (see Figure \ref{fig:brightness_temp} and Table \ref{tab:brightness_temps_lit} in Appendix \ref{sec:Appendix_additional_figures}). The planets are divided into those above (circles) and below (triangles) the expected transition in planet density from rocky to not (just) rocky ($\sim$0.8 in $\rho$/$\rho_e$, the ratio between the density of the planet and the density for the same mass planet given an Earth-like composition; \citealt{Plotnykov2024}). Given the high irradiation and low escape velocity of TOI-561~b (the light blue symbol outlined in dark blue), our JWST/NIRSpec evidence for a thick atmosphere is in strong conflict with the empirical ``cosmic shoreline'' hypothesis. 
}
    \label{fig:cosmic_shoreline}
\end{figure*}

This evidence is corroborated by the planet's anomalously low bulk density ($\rho$/$\rho_e \sim 0.68$, \citealt{Plotnykov2024}), which supports the presence of a volatile envelope, as the planet radius lies 1$\sigma$ above the expected value for a purely rocky planet (pure MgO+SiO$_2$ interior; Figure \ref{fig:cosmic_shoreline}). A volatile-rich envelope around TOI-561~b demonstrates that ultrahot super-Earths can hold onto atmospheric volatiles on Gyr timescales. The ability of TOI-561~b to retain a substantial atmosphere over billions of years is especially surprising, given its low mass and current extreme radiation \citep{Owen2019}. Our finding strengthens the hypothesis that planetary-scale magma oceans can act as an internal volatile sink \citep{Dorn&Lichtenberg2021}, limiting loss to space \citep{Lichtenberg2025}, although more theoretical work is needed to better understand the atmospheric escape of higher-molecular-weight gases (see the initial forays by \citealt{Chatterjee2024,Ji2025}). From the sample of rocky planets with dayside brightness temperature constraints (Figure \ref{fig:brightness_temp}), it appears that planets with irradiation temperatures exceeding $\sim$2000~K are able to replenish volatile envelopes faster than they are lost. These observations call into question the universality of the empirically defined ``cosmic shoreline'' between planets with and without atmospheres, as shown in Figure \ref{fig:cosmic_shoreline}, which uses the very approximate scaling estimate for cumulative X-ray and ultraviolent (XUV) irradiation from \cite{2017ApJ...843..122Z}. 

Pinpointing exactly why TOI-561~b has a thick atmosphere will require further theoretical and observational investigation. Although the explanation for a thick volatile envelope around TOI-561~b could in principle include alternative formation scenarios, such as the late delivery of volatiles \citep{OBrien2018} or late arrival to an ultrashort orbital period \citep{Pu&Lai2019,Schmidt2024}, these scenarios appear unlikely, given the compact architecture and old age of the TOI-561 system. The host star is a thick-disk star that is relatively depleted in iron and enhanced in alpha elements \citep{Weiss2021,Lacedelli2022}, representing a chemically distinct environment from other USP planets and the solar system. This makes TOI-561~b's expected interior abundance ratios of refractory elements (Fe/Mg, Fe/Si) distinct from solar-type compositions \citep{Spaargaren2023}. With comparable spectral data from USPs orbiting host stars with a range of compositions, we can test how iron abundance impacts volatile retention -- that is, whether the volatiles are chemically bound in the molten magma \citep{Bower2022,Lichtenberg2021JGRP} or iron core \citep{2022PSJ.....3..127S,Luo2024} -- opening possibilities to probing magma ocean dynamics \citep{Lichtenberg2021,Meier2023} and the mechanism of volatile accretion \citep{Kite&Schaefer2021,Kimura2022}. The initial volatile endowment \citep{Krijt2023,Lichtenberg2023}, surface temperature \citep{Boukare2022}, and geophysical and geochemical interaction with the interior \citep{Lichtenberg2025} critically affect the ability of low-mass exoplanets to retain and replenish their atmospheres; these factors need to be incorporated into more atmospheric escape calculations. Overall, studying USP planets offers important context for the low-mass exoplanet population and the optimal strategy for characterizing their atmospheres \citep{Redfield2024,Hammond+25}, showing that this will require coupled models of atmospheric escape and geochemical internal evolution. 

\clearpage
\appendix
\renewcommand\thefigure{\thesection.\arabic{figure}} 
\renewcommand\thetable{\thesection.\arabic{table}} 

\setcounter{figure}{0}
\setcounter{table}{0}

\section{Additional Figures and Tables}\label{sec:Appendix_additional_figures}

\begin{figure*}[h]
    \centering
        \includegraphics[width=1\textwidth]{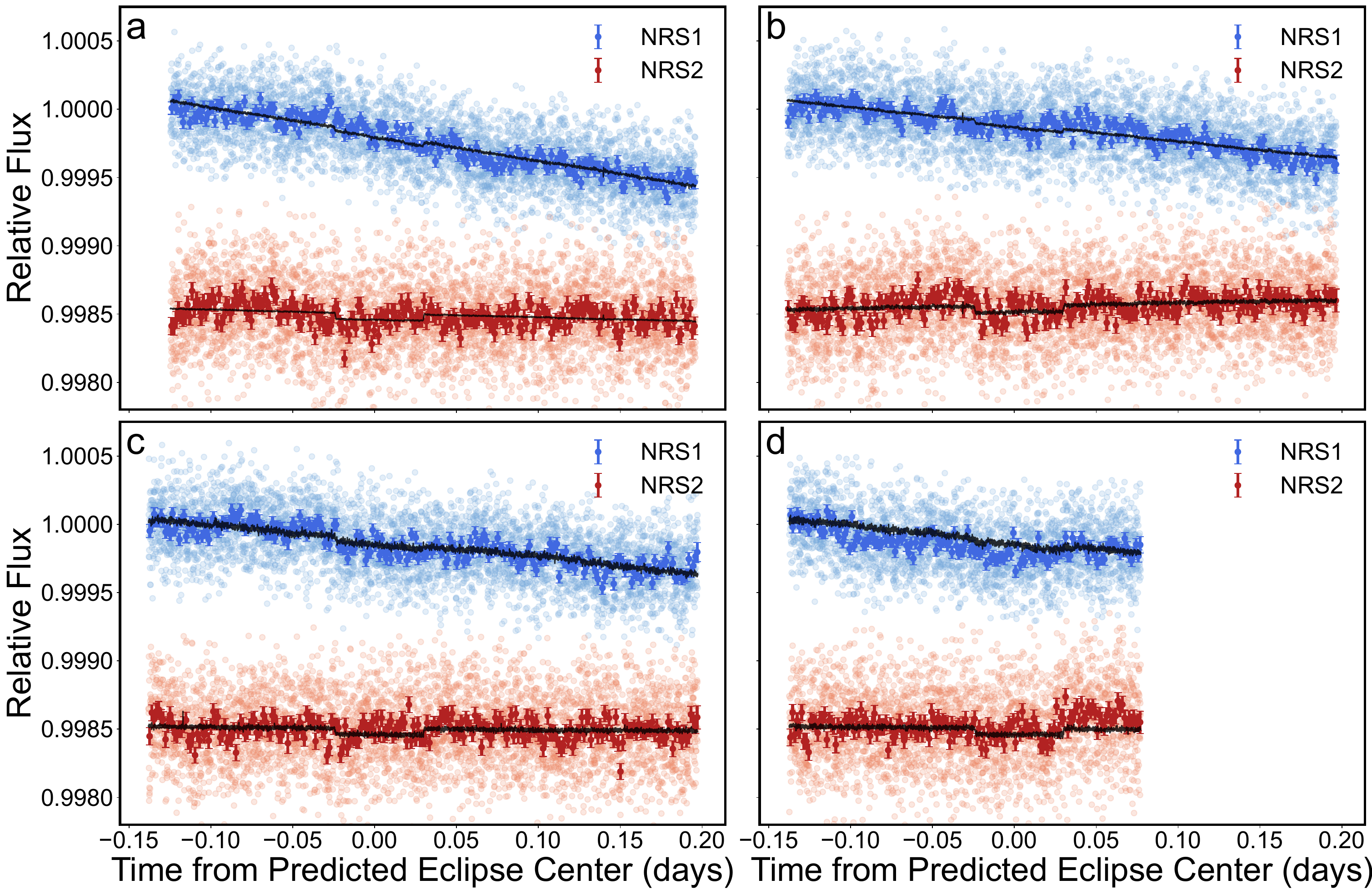}
    \caption{White-light curves of secondary eclipses of TOI-561~b (a-d). The four individual consecutive secondary eclipse white-light curves with best fit models from the joint fits to the \texttt{Eureka!} eclipses. In each panel, NRS1 (2.8627040 to 3.7143560 microns) is in blue and NRS2 (3.8199180 to 5.0627574 microns) is in red. The unbinned points are shown in the background and the two minute binned points as shown with their errors. The best fit models are shown in black.	}
    \label{fig:individual_WLC}
\end{figure*}

\begin{figure*}[h]
    \centering
        \includegraphics[width=1\textwidth]{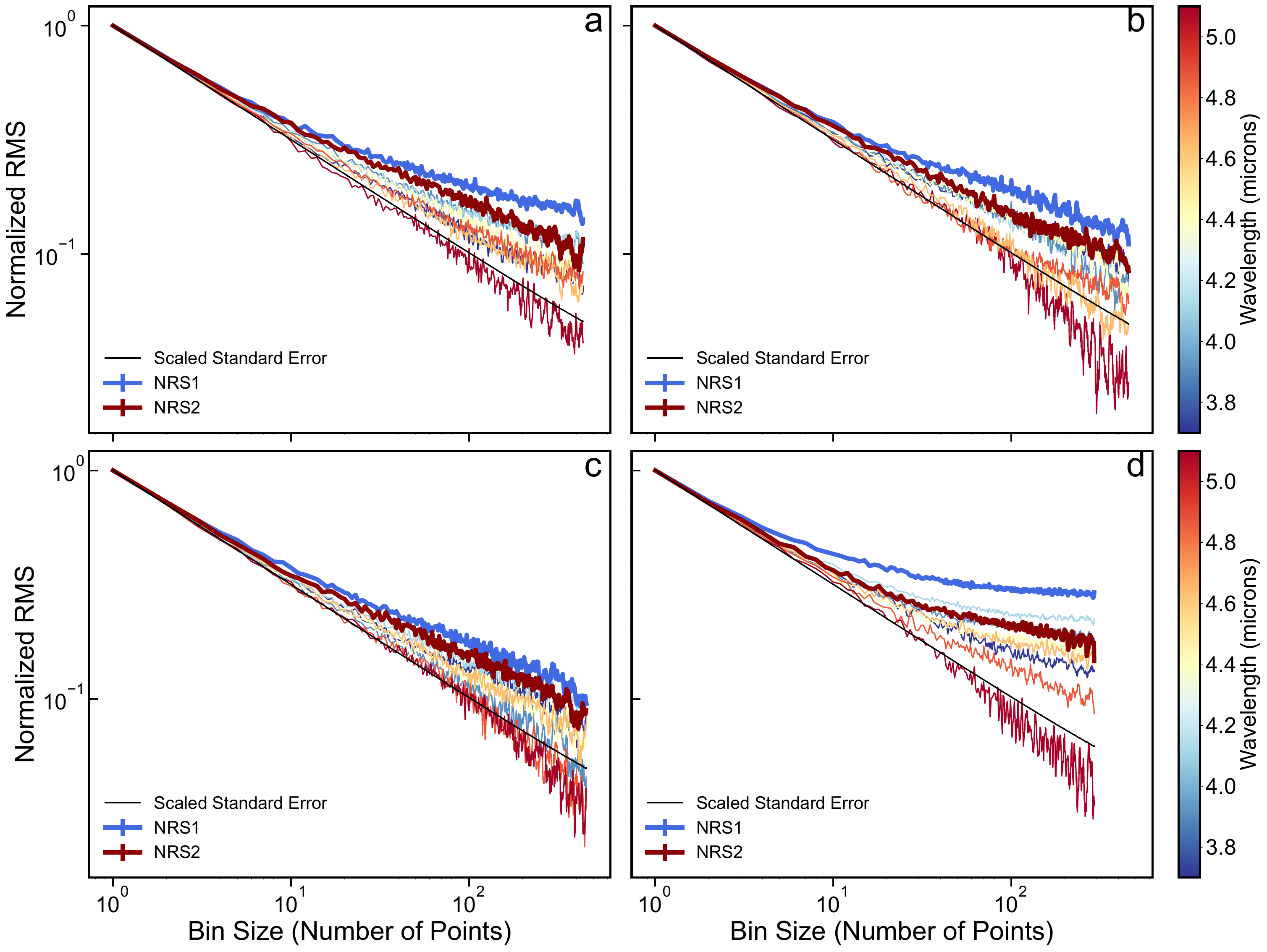}
    \caption{Normalized RMS versus bin size for the \texttt{Eureka!} reduction white-light curves (heavier lines) and spectroscopic light curves (lighter lines). In these plots, NRS1 is in blue and NRS2 is in red, and panels (a-d) show eclipses (1-4). In the absence of time-correlated noise, the actual lines (shown in colors) would follow the $1/\sqrt(n)$ lines (shown in black). There is residual correlated noise primarily in eclipse 4, which has the least amount of out-of-transit baseline. We show the RMS versus bin size for the white-light curves as bold lines and color code the RMS versus bin size of the spectroscopic bins according to the central wavelength. The fourth eclipse shows higher noise, but including it does not actually bias the jointly-fit emission spectrum -- the spectra with and without the fourth eclipse agree at better than 1.5$\sigma$ per point, with a median difference of less than 5 ppm for the \texttt{Eureka!} reduction.
}
    \label{fig:RMS_binsize}
\end{figure*}

\begin{figure*}[h]
    \centering
        \includegraphics[width=0.9\textwidth]{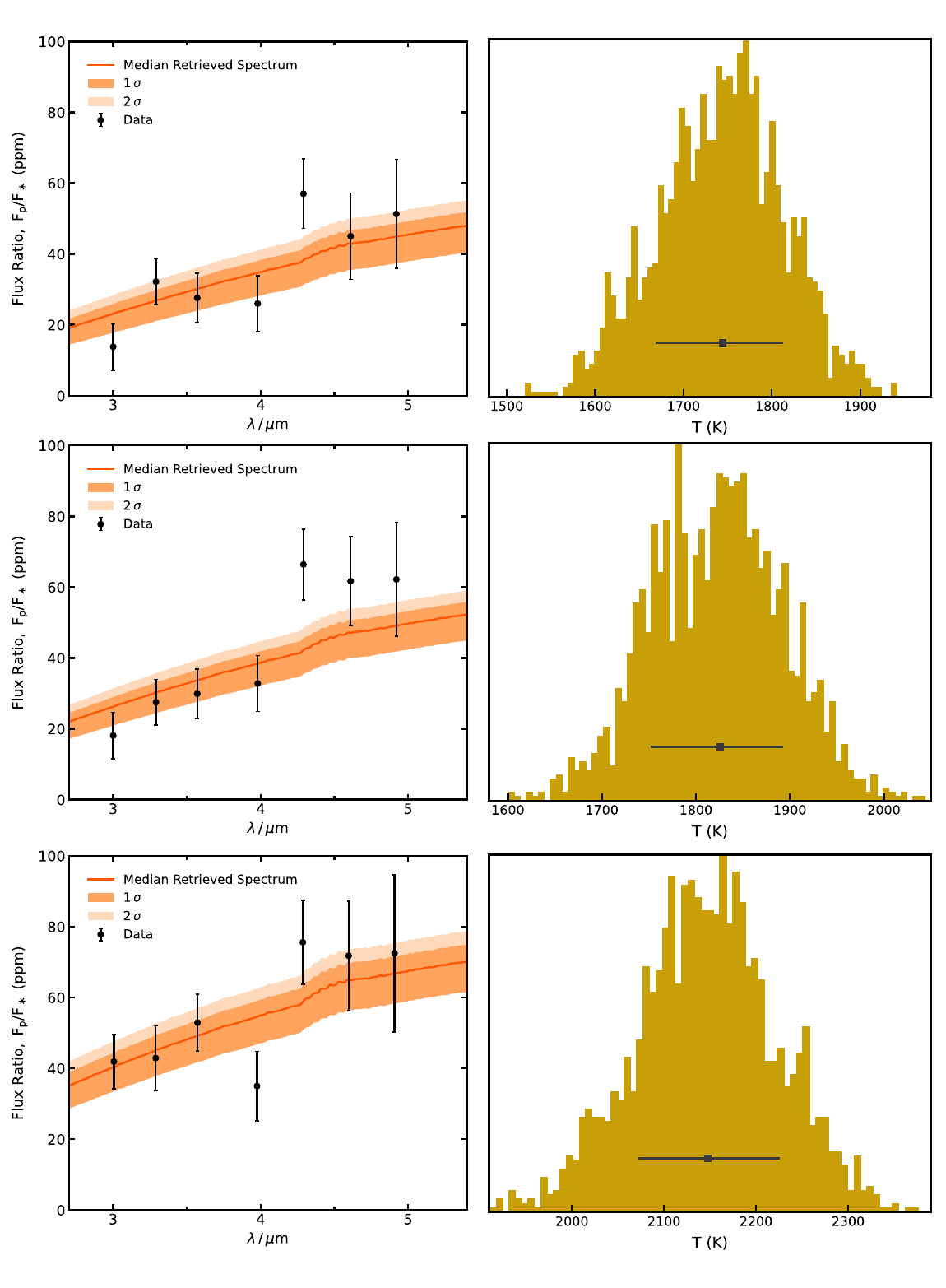}
    \caption{Retrieved blackbody emission spectra (left panels) and corresponding temperature posterior probability distributions (right panels) for the \texttt{Eureka!} (top), \texttt{ExoTiC JEDI 1} (middle) and \texttt{ExoTiC JEDI 2} (bottom) reductions. In the left panels, the dark orange lines show the medium retrieved spectrum, while the dark/light shaded regions show the 1$\sigma$/2$\sigma$ contours, respectively. These retrievals assume a single-parameter blackbody spectrum model.}
    \label{fig:bb_temp}
\end{figure*}

Here we provide supplementary Figures \ref{fig:individual_WLC}, \ref{fig:RMS_binsize}, and \ref{fig:bb_temp} and Tables \ref{tab:sys_params} and \ref{tab:brightness_temps_lit}, which are referenced in the main text.

\begin{deluxetable*}{lll}
\tablecaption{Stellar and planet parameters \label{tab:sys_params} }
\tablewidth{0pt}
\tablehead{\colhead{Parameter}&\colhead{Value}&\colhead{Source}}
\startdata
\textbf{Star} & \\
$R_*$ (R$_\odot$)& 0.843$\pm$0.005 &  1 \\
$T_*$ (K)& 5372$\pm$70 &  2\\
log($g$) & 4.5$\pm$0.12 & 2\\
$[$Fe/H$]$ (dex) & -0.40$\pm$0.12 & 1 \\
$[\alpha$/Fe$]$ (dex) & 0.23$\pm$0.04 &  1\\
Age (Gyr) & 11$^{+2.8}_{-3.0}$ & 1\\
\hline
\textbf{Planet} &  \\
Period (days)& 0.4465689683$^{+0.0000002381}_{-0.0000003152}$  &  3\\
Inclination (deg)&88.178$^{+1.0045}_{-1.0820}$ & 3\\
$a/R_*$ &2.692202$^{+0.017130}_{-0.021185}$ &  3\\
$e$& 0 (fixed) & 3 \\
Midtransit time (BJD) & 2459578.546253$^{+0.000219}_{-0.000205}$ &  3\\
Radius ($R_{\earth}$)& 1.4195$^{+0.0217}_{-0.0224}$ & 3\\
Mass ($M_{\earth}$)& 2.24$\pm$2.0 & 4\\
Density (g~cm$^{-3}$) & 4.3049$^{+0.4411}_{-0.4216}$& 3,4 \\
\enddata 
\tablenotetext{}{Sources: (1) \cite{Lacedelli2022}, (2) \cite{Lacedelli2021}, (3) \cite{Patel2023}, and (4) \cite{Brinkman2023}. }
\end{deluxetable*}

\begin{deluxetable*}{lllll}
\tablecaption{Dayside brightness temperature comparison \label{tab:brightness_temps_lit} }
\tablewidth{0pt}
\tablehead{\colhead{Planet}& \colhead{Instrument/Mode} &\colhead{$T_{\rm{irr}}$ (K)} & \colhead{$T_{b,\rm{day}}$ (K)} & \colhead{Reference for $T_{b,\rm{day}}$}}
\startdata
TRAPPIST-1 c & MIRI/F1500W & 480$\pm$5 & 380$\pm$31 & \cite{Zieba2023} \\
TRAPPIST-1 b & MIRI/F1280W & 562$\pm$5 & 424$\pm$28 & \cite{Ducrot2025}\\
TRAPPIST-1 b & MIRI/F1500W & 562$\pm$5 & 478$\pm$27 & \cite{Ducrot2025}\\
TRAPPIST-1 b & MIRI/F1500W & 562$\pm$5 & 503$\pm$26 & \cite{Greene2023Natur.618...39G}\\
LTT 1445 A b & MIRI/LRS & 600$\pm$30 & 525$\pm$15 & \cite{Wachiraphan2025}\\
GJ 1132 b & MIRI/LRS&  826$\pm$14 & 709$\pm$31 & \cite{Xue2024}\\
LHS 1478 b & MIRI/F1500W & 842$\pm$14 & 491$\pm$102 & \cite{August2025}\\
TOI-1468 b & MIRI/F1500W & 964$\pm$11 & 1024$\pm$78 & \cite{MeierValdes2025}\\
GJ 486 b & MIRI/LRS&  985$\pm$10 & 865$\pm$14 & \cite{2024ApJ...975L..22W} \\
LHS 3844 b & IRAC2 & 1138$\pm$28 & 1040$\pm$40 & \cite{Kreidberg2019} \\
GJ 1252 b & IRAC2 & 1540$\pm$98 & 1410$^{+91}_{-125}$ & \cite{Crossfield2022}\\
TOI-1685 b & G395H/NRS2&  1541$\pm$40 & 1360$\pm$100 & \cite{Luque2025}\\
TOI-1685 b & G395H/NRS1&  1541$\pm$40 & 1520$\pm$140&  \cite{Luque2025}\\
GJ 367 b & MIRI/LRS & 1930$\pm$45 & 1728$\pm$90 & \cite{2024ApJ...961L..44Z} \\
TOI-431 b & IRAC2&  2660$\pm$468 & 1520$\pm$360 & \cite{Monaghan2025}\\
55 cnc e & MIRI/LRS & 2834$\pm$13 & 1796$\pm$88 & \cite{Hu2024} \\
K2-141 b & IRAC2 & 3038$\pm$66 & 2049$\pm$361&  \cite{Zieba2022} \\
TOI-561 b & G395H SPECTRUM & 3274$\pm$44 & 1740$^{+70}_{-80}$ & This work, \texttt{Eureka!}\\
TOI-561 b & G395H SPECTRUM&  3274$\pm$44 & 1830$\pm$70  & This work, \texttt{ExoTiC JEDI 1}\\
TOI-561 b & G395H SPECTRUM&  3274$\pm$44 & 2150$\pm$80  & This work, \texttt{ExoTiC JEDI 2}\\
\enddata 
\tablenotetext{}{These are the values plotted in Figure \ref{fig:brightness_temp}.}
\end{deluxetable*}

\clearpage

\section{Further Estimates of Correlated Noise}\label{sec:Appendix_correlated_noise}

\begin{figure*}[h]
    \centering
        \includegraphics[clip, trim={0 0.3cm 0 0},width=1\textwidth]{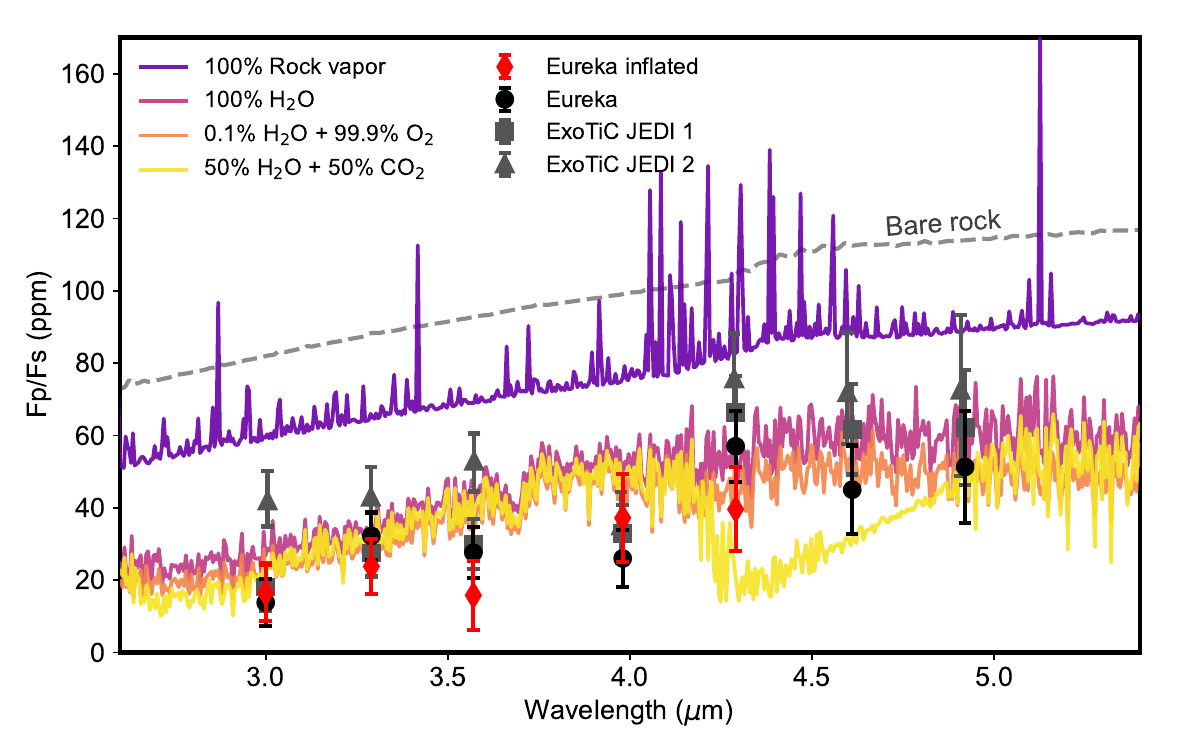}
    \caption{The same as Figure~\ref{fig:spectrum}, but now including the \texttt{Eureka!} reduction$+$fitting showing the estimated increase due to correlated noise, as described in Appendix \ref{sec:Appendix_correlated_noise} (red diamonds and error bars).}
    \label{fig:spectrum_inflated_errors}
\end{figure*}

Our nominal fitting procedure utilized in the \texttt{Eureka!} and \texttt{ExoTiC JEDI 1} results in substantial correlated noise in the white-light-curve residuals (the heavier lines in Figure \ref{fig:RMS_binsize}). While the residual correlated noise in the spectroscopic bins is reduced relative to that in the white-light curves, likely owing to the larger errors in the spectroscopic bins, there is still so-called red noise in these bins (the lighter lines in Figure \ref{fig:RMS_binsize}). To investigate how this red noise impacts our derived constraints, we calculated an inflation factor for our \texttt{Eureka!} reduction errors, according to the method employed by \cite{Pont2006} and explained in detail in \cite{Wallack2021}. In brief, for each spectroscopic bin, we estimate the amount of residual red noise in our fits at characteristic 20 minute bins by determining the factor between our actual noise versus that expected from white noise (i.e., the multiplicative factor that each of the bins are higher than the scaled standard error in Figure \ref{fig:RMS_binsize} at 20 minute bins). We increase our previously-fit error inflation term by that multiplicative factor when it is greater than 1 for all visits of a spectroscopic bin (which it is not for the last two bins of NRS2). We then rerun our fits of the spectroscopic binned light curves with these updated (fixed) error inflation terms (which are added in quadrature to our measured errors) and produce a joint fit spectrum with inflated errors, shown in Figure \ref{fig:spectrum_inflated_errors} as red diamonds. We find that these inflated errors result in reduced chi-square values of 28.2/4.3/2.3/2.9 for the pure-rock-vapor, 100\% H$_2$O, 0.1\% H$_2$O$+$99\% O$_2$, and 50\% H$_2$O$+$50\% CO$_2$ models, respectively. 

This analysis adds further evidence that our overall conclusions are robust to red noise, in addition to the \texttt{ExoTiC JEDI 2} alternative fitting procedure and resulting spectrum.

\begin{acknowledgments}

This work is based on observations made with the NASA/ESA/CSA James Webb Space Telescope. The data were obtained from the Mikulski Archive for Space Telescopes at the Space Telescope Science Institute, which is operated by the Association of Universities for Research in Astronomy, Inc., under NASA contract NAS 5-03127 for JWST. These observations are associated with program 3860. Support for program 3860 was provided by NASA through a grant from the Space Telescope Science Institute, which is operated by the Association of Universities for Research in Astronomy, Inc., under NASA contract NAS5-03127. Support for program 3860 was provided by the Canadian Space Agency under contract 23JWGO2B06. This work has been partially funded by the Natural Sciences and Engineering Research Council of Canada (grant RGPIN-2021-02706). We would like to acknowledge that our work was performed on land traditionally inhabited by the Wendat, the Anishnaabeg, Haudenosaunee, Metis, and the Mississaugas of the New Credit First Nation.

T.L. was supported by the Branco Weiss Foundation, the Netherlands eScience Center (PROTEUS project, NLESC.OEC.2023.017), the Alfred P. Sloan Foundation (AEThER project, G202114194), and NASA's Nexus for Exoplanet System Science research coordination network (Alien Earths project (80NSSC21K0593).

J.T., T.L., A.P., N.W., and R.P. thank the AEThER project, funded by the Alfred P. Sloan Foundation (G202114194), for the opportunity to discuss ideas related to this manuscript. R.P. and A. M. additionally received support from the U.K. Science and Technology Facilities Council consolidated grant ST/W000903/1.

L.D. acknowledges support from the Natural Sciences and Engineering Research Council (NSERC) and the Trottier Family Foundation.

We thank the referee for the thorough review and helpful comments that improved the quality of this manuscript.

Co-author contributions are as follows: S.B. provided independent analysis of the full data set, reinforcing confidence in the results presented here. L.D. performed an independent fit to the \texttt{ExoTiC JEDI} reduction, compiled dayside temperature measurements, contributed to the manuscript writing, and supervised a student. T.L. contributed to the the physical interpretation, manuscript writing, and student funding. R.P contributed to the physical interpretation of the modeling and observational results and student supervision. A.P. performed the theoretical modeling analyses and wrote the modeling sections of the manuscript, including figures. M.P. contributed to the physical interpretation, including interior modeling that went into Figure \ref{fig:cosmic_shoreline}. E.P. contributed to the \texttt{ExoTiC JEDI} data reduction. J.T. designed the observations, contributed to the \texttt{ExoTiC JEDI} data reduction, led the manuscript development, and contributed to the manuscript writing, including figures. N.W. performed the \texttt{Eureka!} reduction and light-curve fitting of both the \texttt{Eureka!} reduction and \texttt{ExoTiC JEDI} reduction, and wrote these sections of the Letter, including figures and appendices. All coauthors contributed to the observing proposal and/or manuscript editing.

\end{acknowledgments}


\bibliography{references}{}
\bibliographystyle{aasjournalv7}

\end{document}